\documentclass[9pt,twocolumn,twoside]{article} % 

\usepackage{hyperref}
\usepackage{graphicx}
\usepackage{color, colortbl}

\title{The human quest for discovering mathematical beauty in the arts\thanks{Originally published in \href{https://www.pnas.org/content/117/44/27073}{PNAS 2020 117 (44) 27073-27075}}.}

\usepackage[natbib=true,backend=bibtex,sorting=none]{biblatex}

\bibliography{info.bib}

\begin{document}

\author{Stefano Balietti}
\date{\footnotesize Mannheim Center for European Social Science Research (MZES), A5, 6, Bauteil A, 68159 Mannheim\\ Alfred-Weber Institute of Economics
Heidelberg University, Bergheimer Str. 58, 69115 Heidelberg\\
 E-mail: {\normalfont stefano.balietti@mzes.uni-mannheim.de}
}

\maketitle

% \subsection*{Format}

% This template is intended for authors writing invited commentaries or perspectives. The format for these article types may vary, but an abstract is required for perspectives. Please be sure to include the title, author line anduthor affiliations, keywords, acknowledgments, and references. Other sections or headings are permitted as needed.

% \subsection*{Manuscript Length}

% Commentary text and references should not exceed 14,000 characters (including spaces). PNAS encourages the use of a single color figure or table since they help summarize the article for scientists outside the immediate field of the paper. 

%  What makes the contributions particularly interesting or important? How do they relate to past work; do they raise new questions or force reinterpretation of previous work? How will the research advance the field? Please avoid jargon and write for a multidisciplinary audience. 

In the words of the twentieth-century British mathematician G.H. Hardy, ``the human function is to 'discover or observe' mathematics'' \citep{livio_golden_ratio_2008}. For centuries, starting from the ancient Greeks, mankind has hunted for beauty and order in arts and in nature. This quest for mathematical beauty has lead to the discovery of recurrent mathematical structures, such as the golden ratio, Fibonacci, and Lucas numbers, whose ubiquitous presence have been tantalizing  the minds of artists and scientists alike. The captivation for this quest comes with high stakes. In fact, art is the definitive expression of human creativity, and its mathematical understanding would deliver us the keys for decoding human culture and its evolution \cite{SigakiE8585}. However, it was not until fairly recently that the scope and the scale of the human quest for mathematical beauty was radically expanded by the simultaneous confluence of three separate innovations. The mass digitization of large art archives, the surge in computational power, and the development of robust statistical methods to capture hidden patterns in vast amounts of data have made it possible to reveal the---otherwise unnoticeable to the human eye---mathematics concealed in large artistic corpora. Starting from its inception, marked by the foundational work by Birkhoff (1933)\cite{birkhoff1933aesthetic}, progress in the broad field of computational aesthetics has reached a scale that would have been unimaginable just a decade ago. The recent expansion is not limited to the visual arts \cite{SigakiE8585} but includes music \cite{10.3389/fnhum.2017.00263}, stories \cite{reagan_emotional_arcs_2016}, language phonology \cite{language3}, humor in jokes \cite{WESTBURY2016141}, and even equations \cite{zeki_math_beauty_2014}; for a comprehensive review see \cite{perc2020beauty}.

\begin{figure*}%[tbhp]
\centering
\includegraphics[width=1\linewidth]{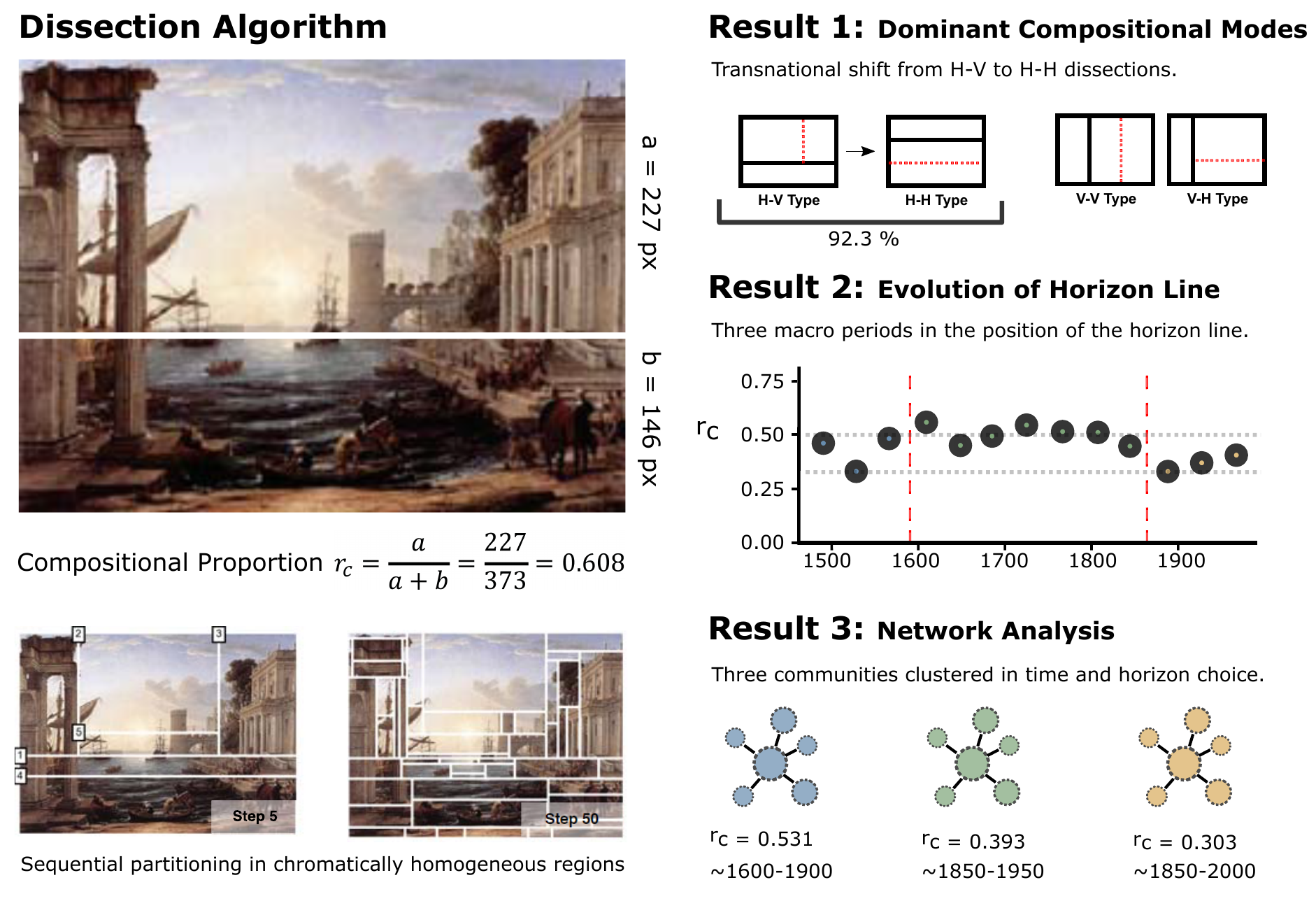}
\caption{Methodology and main results in Lee et
  al. \cite{lee_landscape_proportions_2020}. The vast majority of
  landscape paintings features a first horizontal partition, while the
  direction of the second partition evolved from vertical (H-V) to
  horizontal (H-H); this shift is consistent across individual
  artists' nationalities (Result 1). The ratio of the compositional
  proportion $r_{c}$ in horizontally partitioned paintings denotes the
  height of the horizon line; Lee et
  al. \cite{lee_landscape_proportions_2020} map its progression into
  three macro historical periods (Result 2; graph reconstructed from
  figure 3A). Network analysis reveals the existence of three coherent
  communities of artists clustered in time and in terms of their
  horizon choice (Result 3).  Painting in illustration is ``Seaport
  with the Embarkation of the Queen of of Sheba'' (1648) by Claude
  Lorrain (1604-1682), adapted from Fig. 1 in Lee et. al
  \cite{lee_landscape_proportions_2020}. Painting images credit: The
  National Galley, London.}
\label{fig_lee}
\end{figure*}

In PNAS, Lee et al. \cite{lee_landscape_proportions_2020} extend this quest by looking for statistical signatures of compositional proportions in a quasi-canonical dataset of 14,912 landscape paintings spanning the period from Western Renaissance to Contemporary Art (from 1500 CE to 2000 CE). They use an information-theoretical framework based on the work of Rigau et al. \cite{rigau_info_aesthetics_2008} to mathematically study how painters arrange the colors on the canvas across styles and time (see Fig. \ref{fig_lee}). They implement a computational algorithm that dissects each painting in their dataset into vertical and horizontal regions that are most homogeneous in colors. Their algorithm works sequentially and at each step maximizes the mutual information between colors and regions over all possible partitions in both horizontal and vertical dimensions. As information ``encodes counterfactual knowledge and describes the amount of uncertainty or noise in a system'' \cite{balietti_optimaldesign_2020}, intuitively, gaining  information in this context means becoming more certain that the palettes of the partitioned regions are chromatically distant. Lee et al. \cite{lee_landscape_proportions_2020} validate this approach by comparing the compositional information for abstract and landscape painting, showing that the information gained by early partitions in landscape paintings is markedly higher than in abstract paintings, which show no directional preference.

Lee et al.'s \cite{lee_landscape_proportions_2020} dissection analysis reveals hidden meta patterns of community consensus completely abstracting from considerations of human aesthetics. Yet, the result is a consistent macro history of landscape painting in Western art. What is more, their study offers a precise quantitative understanding of the interconnections between artistic styles, movements, and artists. Are there universal organizing principles that mathematically define artworks across styles and artists? Do these principles differ across nations and cultures? How do they evolve over time? The work of Lee et al. \cite{lee_landscape_proportions_2020} answers all these questions.

At the heart of their analysis, there is a map of the dominant modes of composition of landscape paintings. Based on the direction of the first two dissections, four pairs are possible: horizontal-horizontal (H-H), horizontal-vertical (H-V), vertical-horizontal (V-H), and vertical-vertical (V-V). Because early partitions are the most informative, even this simple categorization is enough to uncover the existence of a smooth transnational shift over time. Initially, the dominant dissection pair was H-V, representing landscapes with at least one large object in the foreground (for instance a building as in Fig. \ref{fig_lee}). However, from the mid-eighteenth century, the ratio of H-H paintings started to surge, rapidly becoming the dominant one in the next century. This result is important in and on itself because it traces a global change in the style and taste for the composition of landscape paintings in the direction of wider horizons with multiple planes in perspective. However, it is even more important because this pattern consistently holds at the level of individual nationalities (as canonically attributed to artists).
%This pattern might partly capture the shift toward linear perspective throughout Europe.

% \paragraph{Result 2: Nationality not as important}
Lee et al. \cite{lee_landscape_proportions_2020} track the evolution of the landmark feature of landscape paintings: the position of the horizon line. They define a measure of compositional proportion $r_c$ as the ratio between the height of the first partition and the total height of the painting (for this analysis they used only paintings with a first horizontal partition, roughly 92.8\% of their total dataset). Over the years, the unfolding of the compositional proportion $r_c$ well encompasses known trends in the history of landscape painting, unveiling three macro periods. The first period is characterized by low values of $r_c$, found mainly in the mid-sixteenth century and exemplified by paintings with large aerial views, such as those by Pieter Bruegel the Elder. Subsequently, the values of $r_c$ gradually increase until reaching a peak at the beginning of the seventeenth century and remaining high throughout the mid-nineteenth century; the grandiose panoramas of romantic painters such as Caspar David Friedrich belong to this second period. In the last stretch, the level of $r_c$ shrank again to lower values; however, the tails of the $r_c$ distributions became more prominent, indicating more variability in a period historically associated with more stylistic diversity. A major contribution of this analysis is that it reveals surprising cross-style similarities: even throughout the Cambrian explosion of styles of the twentieth century, the values of $r_c$ remain confined in a relatively tight interval around the value of 1/3.

% choices cross-correlates with other artists and styles. 
Using network analysis, Lee et al. \cite{lee_landscape_proportions_2020} investigate the horizon placement at the level of individual artists. They construct a compositional similarity network, weighting the links between each pair of artists and styles depending on how similar their distributions of $r_c$ are. After pruning low significance connections, a standard community detection algorithm reveals the existence of three groups of artists, clustered in time and in terms of their horizon choices. The first community is characterized by a high value of $r_c$ (slightly below the middle of the painting) and spans from the seventeenth century until roughly the end of the twentieth century. The second community is characterized by lower $r_c$ values and is concentrated between the end of the nineteenth and the beginning of the twentieth century. Finally, the last community is mainly found in the twentieth century and features artists with lowest values of $r_c$, but also the largest standard deviation. Overall, it is impressive that, absent any meta data about time and style, this analysis manages to reconstruct coherent communities and, what is more, to highlight important bridges between them.

It is worth commenting here on the connection between the computational results by Lee et al. \cite{lee_landscape_proportions_2020}---fruit of the latest advances in digital data processing and of the access to affordable computer power---and a foundational theory in art history known as ``significant form.'' Conceived by art critic Clive Bell \cite{bell_art_1914} in 1914---a time in which the only computers were human \cite{nelson_harvard_computers_2008}---this theory argued that the essence of art lies in ``lines and colors combined in a particular way, certain forms and relations of forms, [that] stir our aesthetic emotions.'' Hence, the aesthetic value of a piece of art is entirely derived from forms and relations that evoke a transcendent artistic response, independently of other kinds of human emotions. In this sense, the work by Lee et al. \cite{lee_landscape_proportions_2020} is a testament to Bell's theory because it makes apparent to the public eye exactly those forms and relations whose knowledge would otherwise be reserved only to trained art critics. Why is this of pivotal importance? We tend to conceive art as accessible to everyone, and to a large extent this is true; however, there still exist numerous examples of topical differences between the expert and popular appreciation for art. According to Semir Zeki, neuroscientist and one of the founding figures of the field of neuroesthetics, in order to appreciate hidden mathematical beauty, we need a brain instrumentally trained for the object of observation. Zeki's research has demonstrated the existence of a single area of the brain that correlates with the experience of beauty for musical and visual arts as well as for abstract concepts such as mathematical equations \cite{zeki_math_beauty_2014}. However, in the case of mathematical equations, there are profound differences between the perception of beauty by trained mathematicians and by lay persons. Possessing a trained brain is the key to decode mathematical beauty. Computational algorithms like the one by Lee et al. \cite{lee_landscape_proportions_2020} can help democratize access to mathematical beauty without degrading its concept, by institutionalizing some of its organizing principles and by tracking their evolution over time. To this extent, one of the major results by Lee et al. \cite{lee_landscape_proportions_2020} is the scaling down of the narratives of insulated national productions and isms, in exchange for a multi-perspective and non-linear macro view of Western art history. This view, albeit familiar to the scholarly literature, has not yet followed suit in library classifications and textbooks, therefore remaining less accessible to the general public.

% In a
%  “rare gift of artistic appreciation.” .

% nevertheless foundational, as it
% 809 reveals the streamlined nature of canonic art history, which is
% 810 a valuable and necessary step preceding all future research that
% 811 is interested in putting this biased canon into question. We
% 812 also believe our study can function as a proper starting point
% 813 to quantitatively investigate principles of composition covering
% 814 broader conceptual groupings (beyond artists and conventional
% 815 style periods) and a broader set of cultures and regions. Our
% 816 methodology is readily applicable to paintings and any other
% 817 2D representational form once an appropriate data set is
% 818 prepared. In further analyses, comparing the characteristics of
% 819 photographs as uploaded into online social media, randomly
% 820 selected from street view panoramas, or procedurally generated
% 821 could be intriguing avenues of study

While computational aesthetics is a research area in active evolution, the emergence of quantifiable and verifiable mathematical principles already bears profound implications for both the near and the far future of humanity. First, they immediately enhance the accountability and objectivity of subjective peer evaluation, which are known to suffer from cognitive biases, and self-serving behavior in high-stake domains \cite{balietti_artex_2016}; for the same reason, they can be used to verify the authenticity of artwork \cite{taylor_authenticating_2007}. Second, and more importantly, they contribute to the creation of a foundational toolset that will allow humans to pass the baton of the quest for mathematical beauty in the arts to our successors: artificial intelligence (AI). Current methods in AI can reproduce works of art in the style of fashionable painters \cite{mazzone_art_2019}---some of those even sold for hundreds of thousands of dollars at auction houses---but they still remain too narrow to grasp even a glimpse of the concept of beauty that humans have \cite{marcus2019rebooting}. Despite the amazing progress in several well-defined domains, we are still far from the creation of a true artificial general intelligence capable of complex causal reasoning and abstraction \cite{goertzel_agi_2007}. For this, some authors have invoked a paradigm shift in AI architecture: from blank-slate end-to-end learning, e.g., deep neural networks, to a modular system made of different components, similar to the hierarchical structure of the human brain 
\cite{marcus2019rebooting}. Information-theoretical algorithms like that of Lee et al. \cite{lee_landscape_proportions_2020}, which elegantly summarize macro patterns of the history of human art, could become part of the ensemble of modules teaching artificial brains how to follow human-inspired principles of compassion and beauty-seeking in the arts, but not only in the arts. 

% 10.1525/mp.2011.29.1.109 jazz improvisation entropy
% 10.3389/fncom.2018.00097 improvisation
% Nakayama_2020

The elegance and tractability of information-theoretical approaches have facilitated their application to a broad variety of contexts in computational aesthetics \cite{perc2020beauty}. The tools of mutual information \cite{lee_landscape_proportions_2020}, statistical surprise \cite{10.3389/fnhum.2017.00263}, and permutation entropy \cite{SigakiE8585} have been used to mould the abstract complexity of art into a quantitative form. However, as Claude E. Shannon, the founding father of information theory, warned us in 1956, ``few exciting words like information, entropy, redundancy, do not solve all our problems'' \cite{shannon_bandwagon_1956}. For instance, the algorithm by Lee et al. \cite{lee_landscape_proportions_2020} performs suboptimally with paintings requiring diagonal partitions (like in ``Landscape on The Mediterranean'' by Paul Cezanne) or when large objects are positioned towards the center of the canvas (like in ``The Babel Tower'' by Pieter Bruegel the Elder); furthermore, late partitions yield relatively low information, even if they may conceal highly relevant historical details---a challenge perhaps best tackled in tandem with modern computer vision methods \cite{simoes_movie_ml_2016}. The work by Lee et al. \cite{lee_landscape_proportions_2020} ``does not solve all our problems'' \cite{shannon_bandwagon_1956}, but it is an excellent starting point and has foundational value. By spelling out the riddle of decoding mathematical beauty through the lenses of geometric proportions, it invites scholars from different disciplines to push the quest for mathematical beauty towards broader categories and deeper understanding.

\printbibliography

\end{document}